\documentclass[10pt]{article}

\usepackage{fullpage}
\usepackage{setspace}
\usepackage{parskip}
\usepackage{titlesec}
\usepackage[section]{placeins}
\usepackage{xcolor}
\usepackage{breakcites}
\usepackage{lineno}
\usepackage{hyphenat}

\PassOptionsToPackage{hyphens}{url}
\usepackage[colorlinks = true,
            linkcolor = blue,
            urlcolor  = blue,
            citecolor = blue,
            anchorcolor = blue]{hyperref}
\usepackage{etoolbox}

\usepackage[numbers]{natbib}

\usepackage{authblk}

\usepackage{graphicx}
\usepackage[space]{grffile}
\usepackage{latexsym}
\usepackage{textcomp}
\usepackage{longtable}
\usepackage{tabulary}
\usepackage{booktabs,array,multirow}
\usepackage{amsfonts,amsmath,amssymb}
\providecommand\citet{\cite}
\providecommand\citep{\cite}

\newif\iflatexml\latexmlfalse

\AtBeginDocument{\DeclareGraphicsExtensions{.pdf,.PDF,.eps,.EPS,.png,.PNG,.tif,.TIF,.jpg,.JPG,.jpeg,.JPEG}}

\usepackage[utf8]{inputenc}
\usepackage[english]{babel}

\usepackage{float}

\begin{document}

\renewenvironment{abstract}
  {{\bfseries\noindent{\abstractname}\par\nobreak}\footnotesize}
  {\bigskip}

\titlespacing{\section}{0pt}{*3}{*1}
\titlespacing{\subsection}{0pt}{*2}{*0.5}
\titlespacing{\subsubsection}{0pt}{*1.5}{0pt}

\title{Avatar Visual Similarity for Social HCI: Increasing Self-Awareness}

\author[1,3,4]{Bernhard Hilpert}%
\author[2]{Claudio Alves da Silva}%
\author[1]{Leon Christidis}%
\author[1]{Chirag Bhuvaneshwara}%
\author[1]{Patrick Gebhard}%
\author[1]{Fabrizio Nunnari}%
\author[1]{Dimitra Tsovaltzi}%

\affil[1]{Affective Computing Group, German Research Center for Artificial Intelligence (DFKI)}%
\affil[2]{Saarland University}%
\affil[3]{LIACS, Universiteit Leiden}%
\affil[4]{Department of Computer Science, Vrije Universiteit Amsterdam}%

\vspace{-1em}

  \date{\today}

\begingroup
\let\center\flushleft
\let\endcenter\endflushleft
\maketitle
\endgroup

\selectlanguage{english}
\begin{abstract}
Self-awareness is a critical factor in social human-human interaction and, hence, in social HCI interaction.
Increasing self-awareness through mirrors or video recordings is common in face-to-face trainings, 
since it influences antecedents of self-awareness like explicit identification and implicit affective identification (affinity). However, increasing self-awareness has been scarcely examined in virtual trainings with virtual avatars, which allow for adjusting the similarity, e.g. to avoid negative effects of self-consciousness \cite{chehayeb2021individual}. Automatic visual similarity in avatars is an open issue related to high costs. It is important to understand which features need to be manipulated and which degree of similarity is necessary for self-awareness to leverage the added value of using avatars for self-awareness. This article examines the relationship between avatar visual similarity and increasing self-awareness in virtual training environments. We define visual similarity based on perceptually important facial features for human-human identification and develop a theory-based methodology to systematically manipulate visual similarity of virtual avatars and support self-awareness. Three personalized versions of virtual avatars with varying degrees of visual similarity to participants were created (weak, medium and strong facial features manipulation). In a within-subject study (N=33), we tested effects of degree of similarity on perceived similarity, explicit identification and implicit affective identification (affinity).
Results show significant differences between the weak similarity manipulation, and both the strong manipulation and the random avatar for all three antecedents of self-awareness. An increasing degree of avatar visual similarity influences antecedents of self-awareness in virtual environments. 
\end{abstract}

\sloppy

\section{Introduction}

{\label{Introduction}}
Self-awareness is a critical factor in social human-human interaction and, hence, in social HCI interaction. 
Being aware of their own emotions and thoughts may allow users to reflect on and later adapt behavior. In traditional training settings, self-awareness has been induced through methods of self-representation such as mirrors or video recordings. However, with fast-paced developments in technology that facilitate newer, more interactive, and immersive virtual training settings (e.g. \cite{bhuvaneshwara2023mithos}), the self-representation methods must be adequately adapted.
Avatars are virtual agents that act as representations of individuals and may be controlled by a human. As such, they can be used as personalized awareness tools to increase self-awareness and are often used in virtual environments such as online games, social media, and, more recently, in virtual worlds \cite{fox2013avatars}. The concept of virtual representation, or the use of avatars as a means of visual representation in virtual environments, has gained significant attention in recent years \cite{downs2019polythetic}. Previous work has investigated the use of digital awareness tools to increase, especially group social awareness \cite{tsovaltzi2017effects, puhl2015long, tsovaltzi2017group, tsovaltzi2014group}. However, the use of avatars to increase self-awareness has been less investigated and has focused on supporting emotion regulation in general settings \cite{schneeberger2018exhail} and in teacher professional development settings \cite{kunter2011professionelle, park2022implicit}. Still, little is known about how avatars can be used to differentially influence self-awareness for different purposes and what the mechanism behind this influence is. Previous research on increasing self-awareness, concentrated on using mirrors \cite{carver1978self} and video recordings \cite{carver1978self, silvia2004self} as external representations of oneself. The assumption here is that the external representation needs to be similar to the individual. An avatar that is visually similar to oneself, might create the same effect for virtual environments. An added value of avatars as external representations is, that they act as "bridges" to virtual worlds in the sense of mixed-reality or XR \cite{kocur2022designing}, which can be used for immersive but adjustable interactions, beyond what is feasible in the real world and at a low cost \cite{ke2016teaching}. They also offer themselves for social interaction training, for instance for developing empathy and conflict resolution strategies, in school contexts and beyond \cite{dalinger2020mixed}. In the same sense, the avatars themeselves can be adjusted to better serve the research or training purposes. One main aspect that has drawn a lot of attention are personalized avatars, a term that refers to the visual similarity to the individual. But what degree of such visual similarity is most beneficial for increasing self-awareness, and how should this similarity be defined?   



There exists conflicting evidence regarding the degree of similarity needed and how similarity should be defined. Some studies show that more similarity is good (e.g., \cite{yee2007proteus}), while other studies and empirical findings (e.g., uncanny valley) show too much similarity might have adversary effects \cite{Stein2017}. A strongly realistic (i.e., similar) avatar can, for example, be experienced as “rather heterogeneous or inharmonious to the surrounding virtual environment” \cite{jo2017impact}. 
In order to examine further the effects of the avatar on a user, we first need to understand which features influence perceived similarity and to be able to define how to manipulate specific facial features of virtual avatars. 
This article therefore investigates, how we can define the degree of similarity to be able to manipulate personalized avatars and increase self-awareness (RQ 1). To this end, it extents and refines a technical methodology presented first in \cite{alves2023visual} to manipulate specific facial features of virtual avatars by the degrees of visual similarity to the user. Further, the article also contributes to defining the preconditions of increasing self-awareness with avatars (namely, perceived similarity as well as explicit and implicit identification) to guide further research on topics involving avatar visual similarity. Subsequently, we test the proposed methodology in a comprehensive user study with regard to its effects on all three antecedents (RQ 2), examining to what extent the systematic manipulation of visual avatar similarity influences perceived similarity (RQ 2.1) as well as explicit (RQ 2.2) and implicit identification (RQ 2.3).
\section{Theoretical Background}

    \subsection{(Virtual) Emotion Regulation Trainings}

{\label{Related Work}}

Social interaction can be very emotionally challenging on a multitude of levels, especially when it comes to conflict of interests and different socio-emotional perspectives. This challenge increases in multi-party and multi-cultural interaction, like modern classrooms. Teachers find it hard and are often insufficiently trained to deal with the integration of different cultural backgrounds and the socio-emotional conflicts that these may give rise to. They need to adequately manage both heterogeneous learning groups with regard to the taught content, and behavioral conflicts in class in order to secure a constructive learning environment and support student development \cite{krause2013arbeitssituation, tsovaltzi2020influence}. This can be very stressful and requires an effective capacity for conflict regulation \cite{duarte2017empathy}, taking into account multiple student perspectives at once, without interrupting the ongoing lesson.
Students are sharp in perceiving subtle emotions and underlying aggression in teacher behavior, and recursive teacher-student discussions disturb classroom conflict resolution \cite{frenzel2021teacher}. Emotion regulation strategies that contribute to a positive classroom climate involve reappraisal, as opposed to suppression \cite{Jiang2016}, and self-compassion, which can support understanding and coping with one’s own and other’s emotions in learning settings \cite{park2022implicit}. However, emotion regulation presupposes self-awareness \cite{Price2018}. Non-effortful implicit emotional awareness (IEA, e.g. self-awareness) and effortful explicit emotional awareness (EEA, e.g. self-compassion) jointly influence higher emotion regulation \cite{gyurak2011explicit}. 
Misinterpreting students' reactions occurs more frequently between different cultural groups, where different modes of communication, especially non-verbal, lead to confusion. When students express in an emotional way that they do not feel integrated into the class, like e.g. “You don't tell me what to do”, teachers may often feel like their authority is being questioned, or that they do not have the class under control. If they are self-aware, they may apply emotion regulation strategies to combat these feelings of inadequacy, threat or shame effectively, and move on to deal with the situation.
Training teachers’ self-awareness and consequently emotion regulation can thus improve subtle (non-)verbal behavior and conflict regulation without superimposing additional unnecessary cognitive effort and stress. However, teacher training in realistic conflict situations and practicing in real classrooms as a unique possibility \cite{krause2013arbeitssituation} pose ethical risks.
Mixed-reality trainings, implementing interactive simulation technology \cite{greenwald2017technology, gebhard2005alma, gebhard2018marssi} and socially interactive agents (SIAs) with positive effects on human emotion regulation \cite{Schneeberger2021Stress, hilpert2021employing} may offer a viable solution. They can provide a safe space to train situational and self-awareness and resulting conflict behavior through realistic virtual interactions \cite{schneeberger2019can}. As they become more and more interactive and immersive, they allow users to interact as if they were in a real classroom, and practice the handling of conflict situations. Further, they offer options for playback and natural social interactive feedback (e.g., \cite{bhuvaneshwara2023mithos}) which can act as awareness tools to make users aware of their own behaviors and others' reactions and induce self-awareness by supporting reflection \cite{Tsovaltzi2017}. Despite these promises, many questions on the psycho-social design of such environments require further investigation in order for these promises to be held. One of these questions is the degree of avatar similarity conducive for self-awareness.   

\subsection{Self-Awareness, Similarity and Avatar Similarity}
    \label{Self-Awareness and Avatar Similarity}
Mirrors \cite{carver1978self} or cameras \cite{silvia2004self} are validated ways to induce self-awareness. Self-related stimuli (e.g. our own face) are more relevant to us than stimuli related to others \cite{bredart2006short}, and the sense of self is inherently linked to one’s own face \cite{porciello2014interpersonal}. In modern VR-training settings where the face is usually covered with the equipment, inducing self-awareness in traditional ways becomes more difficult \cite{petrocchi2017compassion}, as mirrors or video recordings are not possible. Moreover, mixed-reality concepts involve the use of avatars that can cohabit a shared world with virtual agents \cite{Park2021}.
To take this into consideration, recent research introduced the use of personalized avatars. When using visually similar avatars, users felt sufficiently self-represented \cite{vasalou2007constructing} and emotionally attached \cite{hooi2012being} to the stimuli and these avatars could increase the levels of self-awareness in users \cite{kang2020my}.
Avatars are virtual bodies or vehicles that users engage in, in order to interact with a virtual environment \cite{downs2019polythetic}. In addition, they have been conceptualized as symbolic representations of the face for computer-mediated communication \cite{donath2001mediated}.
Similarity between the avatar and the person can be achieved in a variety of ways. Some examples include matching visual characteristics such as hair and clothes \cite{birk2016fostering, jo2017impact}, and matching or mismatching weight \cite{pena2016see}. The specific similarity between the user's and, respectively, avatar's face (subsequently called visual similarity) has been of special importance for the increase of self-presence \cite{aymerich2014relationship}, body ownership, and even identification with the avatar \cite{jo2017impact, suh2011if}. 
Similarity can be either objective, when considering, for example, degrees of observable similarities \cite{rogers1970homophily}, or subjective, referring to the perceived similarity between two people. 
Perceived similarity was found to have a higher impact on predicting self-awareness than observable similarity (i.e. similarity in measurable terms) in some studies (e.g. in \cite{vasalou2009me}).
However, there is scarce literature on similarity for visually detailed and humanoid avatars. Moreover, most studies (e.g. \cite{vasalou2007constructing, aymerich2014relationship}) examine the effects of comparing personalized versus a generic avatar. To the best of our knowledge, there is no previous work on varying the degree of visual facial similarity systematically and testing the optimal degree of similarity to increase self-awareness. 

To address this gap, we propose using avatars as visual representations of humans which offer the possibility to define a systematic manipulation and investigate visually detailed humanoid avatars as a possibility to systematically manipulate their facial similarity. In section, \ref{Methodology}, we present a methodology based on defining facial characteristics and how to embed their manipulation in the whole face. Given the effects found in past literature presented above, we hypothesize, that our new method of manipulation of visual similarity will induce a main effect of avatar-person similarity manipulation on perceived similarity, compared to a random avatar. 

\textbf{Hypothesis 1:} \textit{The avatar will be perceived as most similar in the condition, where the facial features were not systematically manipulated and gradually less similar with an increasing degree of manipulation.}

As a next step in examining the effects of visual avatar similarity on self-awareness, we define the relationship between an individual and their virtual representation or online persona as a sense of identification.
Cohen \cite{cohen2001defining} suggests that explicit identification is a two-stage process, with the first stage being the selection of relevant visual features and the second stage being the comparison of those features to stored representations in memory. This process allows the recognition of familiar objects.
Over a set of carefully constructed studies, \cite{vasalou2007constructing} found a significant correlation between avatar-person similarity and self-awareness, attributed to triggering the same processes of self-representation as a mirror would.
This type of identification occurs because of a perception that characteristics and values are shared and is “a process or state of seeing oneself as similar to, the same as, or fused with another object or person” \cite{downs2019polythetic}. 
Facial avatar-person similarity has been shown to lead to a higher sense of self-presence \cite{aymerich2014relationship} and self-identification \cite{suh2011if}. Thus, given the proposed manipulation of visual avatar-person similarity, we hypothesize, that there will be a main effect of avatar-person similarity manipulation on explicit identification, compared to a random avatar. 

\textbf{Hypothesis 2:} \textit{There will be higher explicit identification with the avatar in the condition, where the facial features were not systematically manipulated, gradually declining with an increasing degree of manipulation.}



Visual similarity is also a key aspect of implicit identification, as it involves the ability to recognize objects that are similar in appearance, even if they are not identical, and influences the intention to use the avatar as well as emotional attachment \cite{suh2011if}.
People identify with their in-game avatars if they have a positive attitude towards them \cite{rahill2021effects} and find the avatar characteristics important to themselves \cite{mayhew2010measuring}.
Subsequently, visual similarity has been shown to increase avatar appreciation (enjoyment, comfort, helpfulness) in users \cite{cui2009exercising}.
They tend to develop a sense of affinity towards their avatar, which can lead to increased self-disclosure and more authentic interactions online \cite{yee2006psychology}.
Affinity can be described as natural liking “driven by subconscious processes that are beyond conscious control” \cite{seymour2021have}. Affinity can also be related to the degree of attraction or similarity between two or more entities. 
There is a significant effect of perceived visual similarity between an avatar and the user on self-awareness \cite{hooi2012being, hooi2014avatar}, which can be attributed to the personalized avatar triggering emotions, beliefs and attitudes like affinity that facilitated identification. Thus, given the proposed manipulation of visual avatar-person similarity, we hypothesize, that there will be a main effect of avatar-person similarity manipulation on implicit affective identification, operationalized as affinity, compared to a random avatar.

\textbf{Hypothesis 3:} \textit{There will be higher implicit affective identification in the condition, where the facial features were not systematically manipulated, gradually declining with an increasing degree of manipulation.}


In conclusion, visual avatar-person similarity that influences both perceived similarity, explicit self-identification as well as implicit affective identification with the avatar and could be used to elicit self-awareness in human users in sophisticated interactive virtual training environments.


\section{A Theory-Based Methodology}
    \label{Methodology}
In this section, the theory-based development of the methodology to systematically vary visual avatar-person similarity is described. 

\subsection{Perceptual Sensitivity and the Selection of Facial Features}
In cognitive science literature, a subset of critical facial features with high perceptual sensitivity (PS) has been identified, which is described as the sensitivity to detect feature differences across faces and that are crucial for determining the identity of faces (i.e. face recognition across different images of the same face) \cite{abudarham2016reverse}. In a set of studies, facial features were adjusted in a systematic and quantitative manner within a so-called "face-space", and perceptual effects of these adjustments were measured to determine a subset of critical features for which participants were found to show high perceptual sensitivity \cite{abudarham2016reverse}. However, their research manipulated specifically static 2D images.
In our approach to transfer these findings to dynamic avatars, the following five critical facial features were selected, based on \cite{abudarham2016reverse}: Chin shape, jaw width, eye shape, lip thickness, and nose shape. 
Beyond their importance for identification, the selection of these five features for the process of avatar-person similarity manipulation was based on their technical feasibility to manipulate them in the open-access operationalization of the methodology discussed in Section \ref{avatar creation}.

 \subsection{Avatar Creation and Modification}
    \label{avatar creation}
In order to ensure the replicability of this work, we aimed to use exclusively openly accessible software tools to create the different degrees of avatar-person similarity manipulation. 	

Previous studies applied different techniques to achieve different levels of avatar-person similarity. \cite{jo2017impact} created avatars comprising a 3D mesh (a structural construction of a polygon-based 3D model), reconstructed from real imagery of participants, towards a cartoon-like virtual character-based avatar created by a 3D artist. Other studies used a 3D scan of the participant's faces to create highly similar avatar faces \cite{suh2011if}, avatars with a face that was modeled by a 3D artist after photographs \cite{aymerich2014relationship}, and online construction tools embedded in applications or games such as Yahoo! \cite{vasalou2007constructing} and Second-Life \cite{hooi2012being}.
The present study was concerned with a way to manipulate the avatar-self similarity by applying a scale on the same continuum, i.e. using the same source of a most similar avatar and afterwards being able to vary from that most similar avatar in a scale (e.g. see Table \ref{tab:deg}), manipulating PS effectively. Thus, a technical set-up that allowed for continuous manipulation of each of the 5 selected high-PS facial features was selected.

To comply with the research goal of creating visually detailed humanoid avatars, we selected existing software that allowed us to generate a realistic (i.e. high visual fidelity) 3D avatar from participant's photos, and then vary the appearance systematically according to differing degrees of similarity.
Our selection criteria included a) the possibility to manipulate specific facial features on a continuous scale, b) the possibility to animate the created avatars (e.g. create short videos of facial movements) and c) a pragmatic approach to only use tools, that are open-access, require minimal technical knowledge and a reasonable amount of time and effort to create multiple avatars for each participant in order to make our methodology applicable for other research teams as well. 
After considering a range of available software tools, \emph{FaceGen Modeller Demo} (facegen.com/modeller.htm) and \emph{MetaHuman Creator} (metahuman.unrealengine.com) were selected to be used in this study. Other tools were considered (e.g. in3D or MeInGame), but subsequently excluded, because they did not fit the criteria above.

\emph{FaceGen Modeller Demo} (subsequently called FaceGen) is a desktop software to generate 3D heads from one or more pictures. The ideal results are obtained from inputting three images (front, left, and right of the face). FaceGen allows modifying 150+ parameters such as demographics (age, gender, racial group), shapes, colours, textures and individual facial features based on slide menus.
Some of the limitations in regard to this study include that it only generates a 3D model of a face (not including hair) and does not include animation capabilities.

To compensate, we also used \emph{MetaHuman Creator}, which is a free cloud-base app developed by Epic Games to create 3D high-fidelity virtual avatars (including hair). The creation of avatars does not require artistic modeling skills, since its 3D creation capabilities are also based on slide menus and interface controls. 
Another advantage of MetaHuman for this scenario of the research is that it allows exporting the 3D avatar to animation tools, which can be used to personalize facial animations including eye tracking.

\subsection{Manipulation of Avatar-Person Similarity}
    \label{Manipulation}
Besides the challenge of creating a manipulable avatar most similar to a person, this research faced the challenge of defining a methodology to vary the degree of similarity between the avatar and the user. As elaborated in (Section \ref{Self-Awareness and Avatar Similarity}), avatar-person similarity has usually been manipulated in the model of low similarity being a generic avatar that might simply match gender, race and ethnicity, to high similarity by matching clothes, or even the participant's faces (e.g. \cite{yee2021facial}).
However, with fast advancements in animation technology \cite{basak20223d} and the rise of platforms for their use like e.g. the Metaverse (e.g., \cite{salagean2023meeting, kim2023avatar}), creating "realistic" personalized avatars and subsequently their similarity becomes relative to the technological progress. 
Even more importantly, as this research aims to better understand on a process level how visual similarity affects self-awareness and its antecedents in interactive virtual trainings, we aimed for a more fine-grained analysis of visual similarity manipulation beyond a simple distinction between personalized vs. non-personalized.

To implement this, a scale was designed to guide the definition of the conditions for creating degrees of similarity variations (i.e. a visual similarity scale, inspired by approaches taken by \cite{park2019understanding}) for this research study that served as the baseline for avatar-person similarity manipulation (see Table \ref{tab:deg}). This approach has the advantage that it can be continuously adapted to the requirements of other scenarios and generalized to future developments and improvements in avatar creation technology, while still keeping the similarity manipulation standardized and scientific.

\begin{table}
  \caption{Degrees of Similarity Manipulation}
  \label{tab:deg}
  \begin{tabular}{ccp{13cm}}
    \toprule
    \textit{S}&\textit{M}&Description\\
    \midrule
    High  & 0\% & Avatar face modeled from participants' pictures (i.e. highly similar with 0\% of manipulation of facial features)\\
    Medium  & 50\% & Avatar face modeled from participant's pictures, then applied a percentage of manipulation (50\%) on selected facial features (chin shape, jaw width, eye shape, lip thickness, nose shape).\\
    Low  & 100\% & Avatar face modeled from participant's pictures then applied a percentage of manipulation (100\%) on selected facial features (see above).\\
  \bottomrule
\multicolumn{3}{l}{\footnotesize \textit{S} = Degree of Similarity, \textit{M} = Degree of Manipulation}\\
\end{tabular}
\end{table}

\begin{figure}[h]
\centering
  \includegraphics[width=0.5\textwidth]{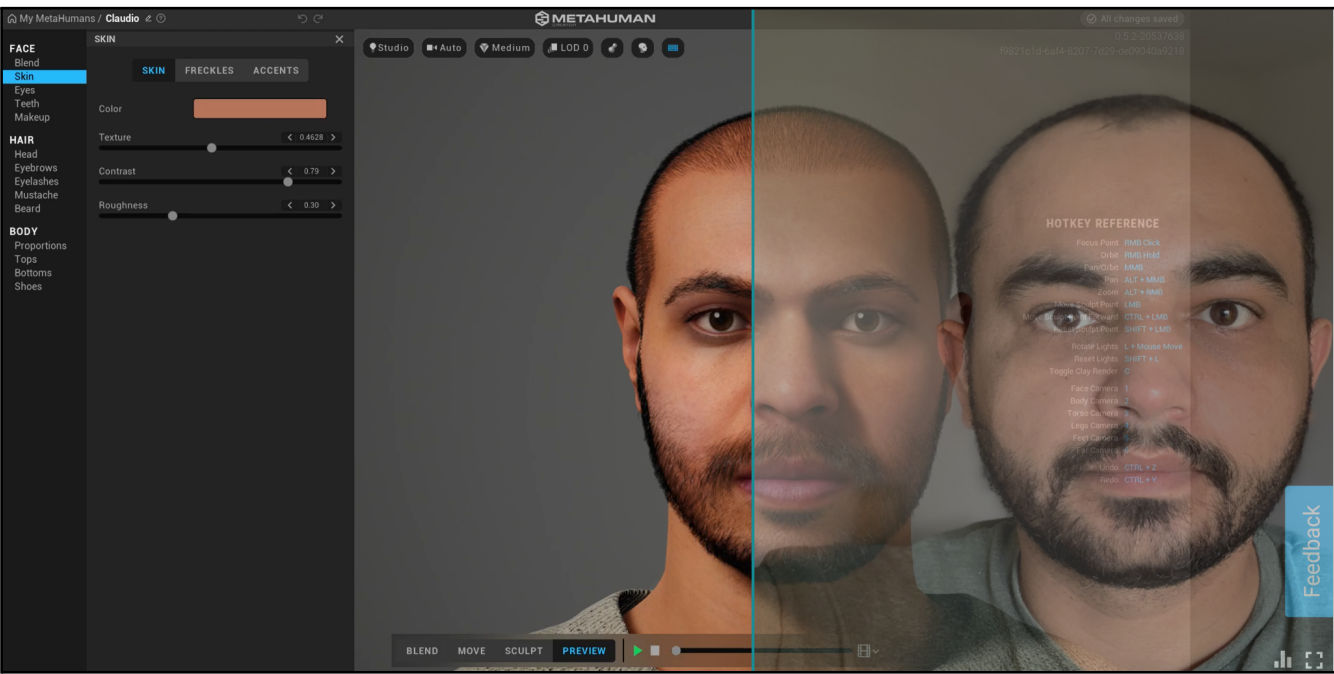}
  \caption{MetaHuman interface and a picture overlay demonstrating the manual creation process}
  \label{fig:Overlay}
\end{figure}

To generate the most similar facial avatar stimuli (0\%-condition), we use MetaHuman Creator, by overlaying a front face photo on the MetaHuman Creator surface and manually adjusting the controls until a satisfactory outcome is reached (see Figure \ref{fig:Overlay}).

\begin{figure}[h]
\centering
  \includegraphics[width=0.5\textwidth]{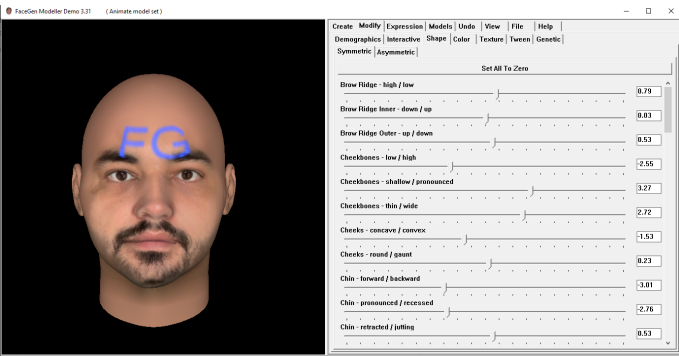}
  \caption{FaceGen interface}
  \label{fig:Facegen}
\end{figure}

For the facial manipulation of the 50\%- and 100\%-conditions, we use FaceGen first, which generates 3D faces from photos and offers modifiers in the form of sliders. FaceGen allows for the distinct manipulation of the facial features on a scale from -10 to +10, with the tool automatically determining the participant’s facial features' scale-value based on their pictures (see Figure \ref{fig:Facegen} for reference).

\begin{figure}[h]
\centering
  \includegraphics[width=0.5\textwidth]{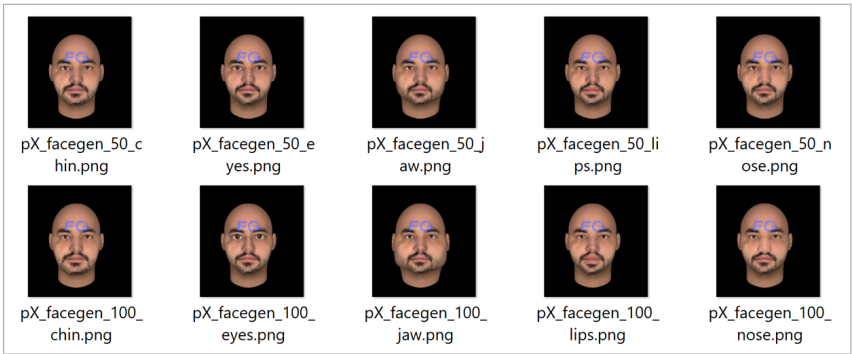}
  \caption{Samples of FaceGen facial manipulation parameters and outcomes}
  \label{fig:Facegen export}
\end{figure}

After FaceGen attributes the scale value for each feature based on the participant’s pictures (0\% condition), the facial variations are then calculated for each feature -  (chin shape, jaw width, eye shape, lip thickness, nose shape) and respective level of manipulation (50\% and 100\%). 
The direction of the similarity degree manipulation in FaceGen (towards -10 or +10) was defined as the direction of the longest distance from the scale value automatically determined by FaceGen upon upload of the photos. For instance, if a face analyzed by FaceGen (0\% condition) points to a chin shape feature having the value of 0.58 in its scale (from -10 to +10), the variations would be towards -10, which is the longest distance from 0.58. Using this example, the manipulation of the chin shape for the 50\% condition would result in updating the chin shape scale value from 0.58 to -4.71 (absolute distance from 0.58 to -10 divided by two minus 0.58), and for the 100\% condition to -10, which is the longest possible distance from the original value, 0.58. An exception is made for the lip thickness feature, which was defined to always move in the positive direction (+10) to avoid deformations on the face that would be impossible to reproduce on MetaHuman (i.e. unrealistic thinner lips sizes). 
Each facial feature is manipulated individually on FaceGen, and each resulting image is exported. This represents a FaceGen export of 10 images per participant (see Figure \ref{fig:Facegen export} for reference).

\begin{figure}[h]
\centering
  \includegraphics[width=0.5\textwidth]{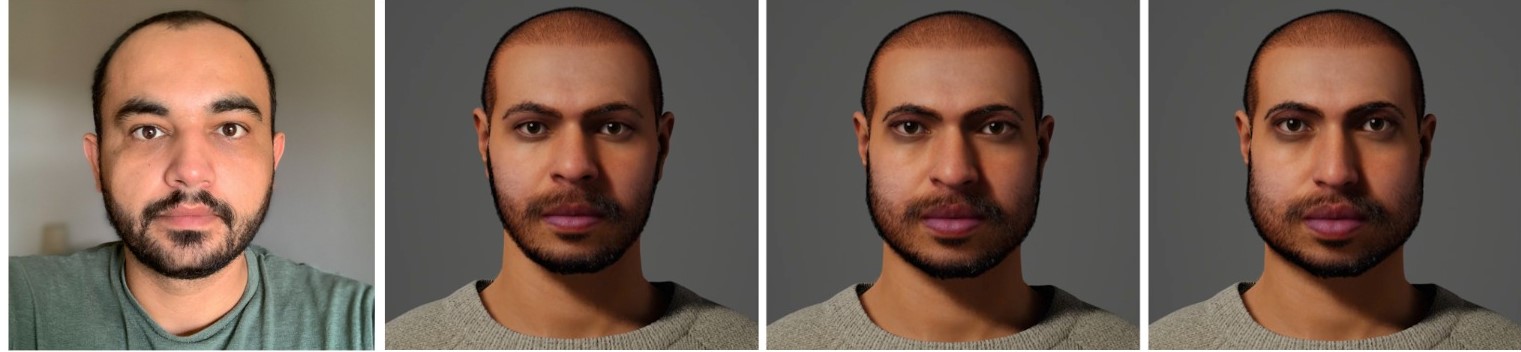}
  \caption{Original photo and the 0\%-, 50\%- and 100\%- variations}
  \label{fig:Conditions}
\end{figure}

As the last step, the first MetaHuman condition made from the participant’s pictures (0\%) is triplicated and each facial feature manipulation in MetaHuman for the 50\% and 100\% conditions is reproduced, following the same process from the first condition to generate the variations by overlaying the exported FaceGen images. Figure \ref{fig:Conditions} illustrates the output of the different conditions from MetaHuman.

\section{User Study}
    \label{Study}

In a first user study, the effects of the varying degrees of visual similarity manipulation on the antecedents of self-awareness were examined. Following the recommendations for open research practices \cite{open2015estimating, wessler2021empirical}, all materials (coding plan, data and analysis code) can be found on OSF\footnote{\url{https://osf.io/zfbjx/?view_only=cbec58b6ab38428aba43c778b511d97c}}. (Note: Due to anonymization, the data and analysis will be made public after the review process, as it contains information that would reveal the authors' nationality.)

\subsection{Participants}

The study was conducted with voluntary participants recruited via various online platforms. Participants were required to be at least 18 years old, have access to a computer/mobile device with a camera, and have a minimum level of English proficiency of B2 according to the Common European Reference Framework for Languages (CEFR; Council of Europe 2001). 
Of the total of \textit{N} = 88 participants, \textit{n} = 55 were excluded due to incomplete responses or not responding to the second email (1).
The final sample of \textit{n} = 33 consisted of \textit{n} = 13 males and \textit{n} = 20 females between 18 and 40 years (\textit{M} = 27.6 years, \textit{SD} = 4.74 years).
We offered participants who completed the study the option of receiving a short video of their personalized avatar and participating in a contest to win Amazon vouchers.

\subsection{Study Design and Procedure}
A within-subjects study was conducted fully online and used self-administered questionnaires delivered via an online survey platform (limesurvey.org). 
In order to systematically test hypotheses 1-3, we used the methodology detailed in Section \ref{Methodology}, to systematically vary the degree of similarity.
Thus, the study consisted of four conditions: high, medium, and low similarity (see Table \ref{tab:deg}) plus one control group (CG).
The study was divided into two sessions, with an interval of 24 to 72 hours between them that corresponds to the time considered to generate the personalized avatars. The main dependent variables were perceived avatar-person similarity, explicit cognitive identification as well as implicit-emotional identification with the avatar.

In the first session, participants were asked to review introductory information about the study and sign the consent form and data protection. Participants confirmed their eligibility by self-reporting their level of English proficiency. If participants reported a proficiency level below the required level, they visualized a message that thanked them for participating and ending the survey. 
Participants with the required proficiency level preceded the survey by filling out a questionnaire that included data collection such as e-mail address \footnote{E-mail addresses were saved in a separate dataset. The merging was done using a unique alphanumeric token. All data that could be used to identify participants were deleted after merging was complete.}, three facial images (as required by FaceGen) and demographic data.

Once the data of the participants from the first survey was received, the images were used to create personalized avatars to varying degrees of similarity to the faces of the participants, as described in Section \ref{Manipulation}. 
Once the avatars were created in MetaHuman, a 12 second-long GIF was created recording the predefined idle facial animation from each of the three conditions (0\%, 50\% and 100\%).
The idle facial animation contained simple facial movements of a natural human look.

As the next step, we set up the second survey by adding the recordings of each personalized avatar condition in the form of animated images (GIF). The conditions and instruments were presented to the participants in random order. The previous same-gender participant’s 0\% avatars were used as a control condition. The first two participants were presented with control-condition a male and female avatar was created ahead from two volunteers' pictures.

\subsection{Measurements and Instruments}
The instruments for collecting the data described in the study design section were the following.
		
\emph{Demographics:}
Participants were asked age and gender, level of formal education, the field and semester of study / current profession. In addition, we collect participants’ affinity with technology, attachment style, and
previous experiences with avatars.

\emph{Perceived facial similarity between avatar and user:} The subscale 'physical similarity' from the polythetic identification model \cite{downs2019polythetic}, containing five items, was used to measure physical similarity (Hypothesis 1).
	
\emph{Explicit cognitive Identification with the Avatar:}
No existing scale was found to measure explicit identification (Hypothesis 2). The self-created items were used on a seven-point Likert scale
, which includes a question based on the definition of identification by \cite{downs2019polythetic}, 'If I look at the GIF, it feels like I am this avatar', and two questions from \cite{van2010player}, 'I identify with this avatar' and 'I feel represented by this avatar'.

\emph{Implicit-emotional Identification with the Avatar:}
 Implicit-emotional Identification (Hypothesis 3) was operationalized as the affinity towards the avatar. We used the affinity scale from \cite{seymour2021have} which contains five items on a seven-point Likert scale.

\emph{Perceived Realism of the Avatar (Manipulation Check)}:
We use a single item based on \cite{jo2017impact}: 'How realistic is the avatar?' on a seven-point Likert scale.

\subsection{Data Analysis}
    \label{Data Analysis}
Prior to conducting the analysis, we assessed the presence of potential outliers, identifying values that exceed Q3 + 1.5xIQR or fall below Q1 - 1.5xIQR as outliers. A total of \emph{n}=5 outliers were detected. On examination, the scores of these outliers were found to be within the range of plausible data, with no indication of measurement errors or inattention by the participants. Furthermore, analyses conducted using both raw data and data excluding outliers yielded similar results. Consequently, we opted to retain the outliers and present findings based on the complete data set.

We hypothesized that for all three tested constructs, there would be a significant main effect for condition, with the most similar (0\% condition) scoring highest for perceived similarity, explicit identification and affinity. A repeated-measures ANOVA was used to examine the effect of condition (0\%-, 50\%-, 100\%-Manipulation, CG) on each of the dependent variables individually.
Shapiro-Test for normality and Mauchly’s Test for sphericity both indicated violations.
However, ANOVAs are rather robust against normality and sphericity violations, see \cite{ito19807}. Upon closer examination of the data, we decided to keep the ANOVA as our main analysis method. In order to ensure, the final reported result would not be biased by either the small sample size or the violations in the preconditions, we performed an additional ANOVA with bootstrapping as well as a Friedman test for non-parametric data. Since these analyses showed identical effects, we report the ANOVA results here. However, all analyses and results can be found in the additional online material of this article.
Moreover, the reported results were corrected using the Greenhouse-Geisser method. Further, we performed post-hoc t-tests to examine the differences between each condition more closely (Bonferroni-Holm corrections were applied). Results are reported individually per construct.

As a manipulation check, we tested the perceived realism of the avatars with one item. While in the 0\%-, 50\% and the CG it was perceived as rather realistic, there was a significant main effect for realism with the 100\%-condition having the lowest mean, indicating, that the avatar was seen as significantly less realistic. 
 
Naturally, in a highly personalized (small sample-size) study like this, individual differences and attitudes of participants may play a role. Since the method and the results of this study will serve as the fundament for a subsequent study series, we additionally opted for an in-depth explorative analysis. Firstly, we examined the correlations between all three dependent measures and gender, realism, experience with avatars, affinity with technology and attachment style. In a second step, we performed several ANCOVAs to examine possible confounding effects of the demographic variables.
Data analyses were conducted with R version 4.2.3.

\subsection{Results}
  
The aim of this research in the context of data collection for the study was to validate the applicability of the methodology developed and test its effects on selected antecedents of self-awareness. 
We hypothesized that for all three tested constructs, there would be a significant main effect for condition, with the most similar (0\% condition) scoring highest for perceived similarity, explicit identification and affinity (implicit identification). 


All means and standard deviations can be found in Table \ref{correlations}.

\subsubsection{Perceived Similarity}
\begin{figure}[h]
\centering
  \includegraphics[width=0.5\textwidth]{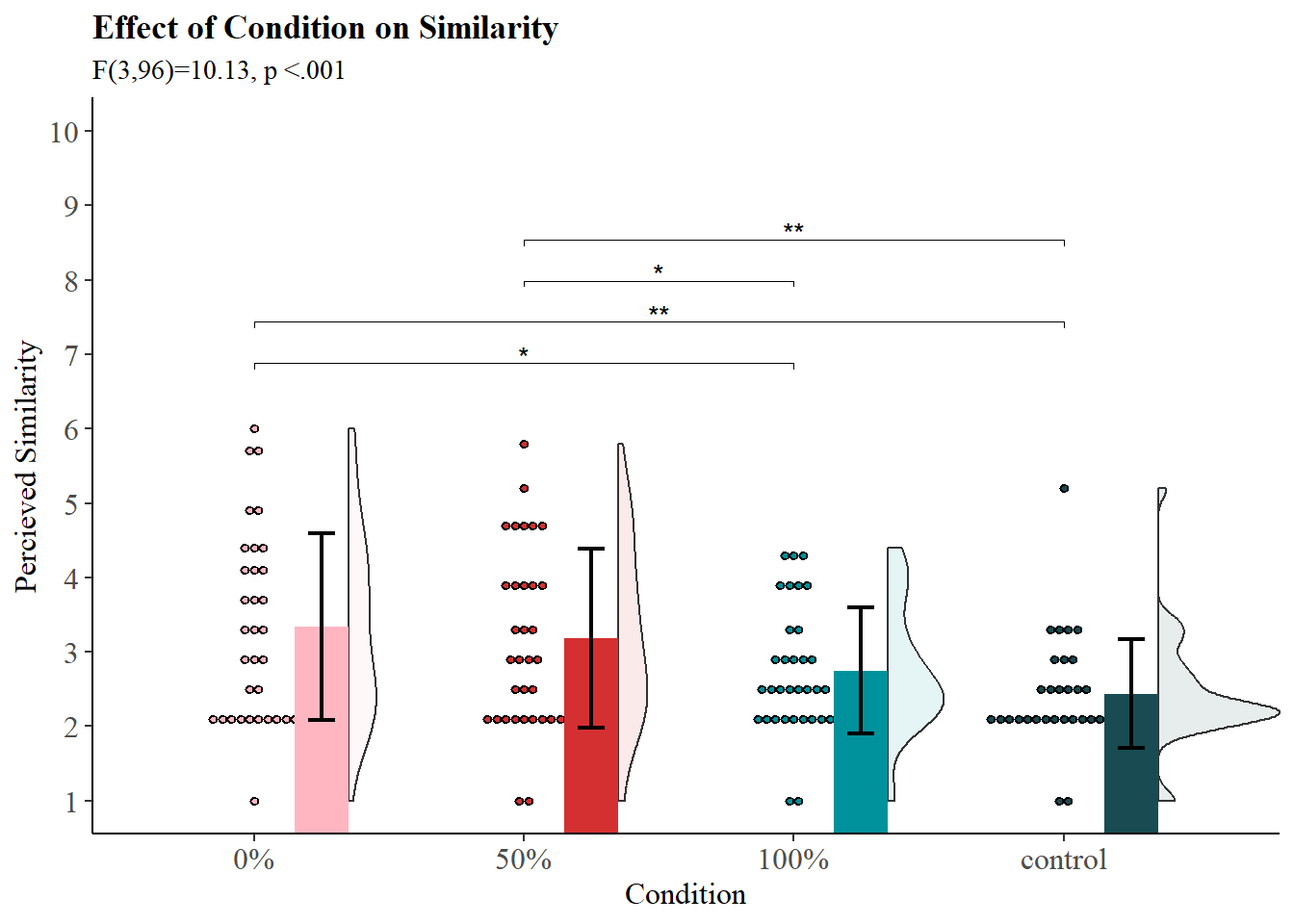}
  \caption{Effects of similarity manipulation on perceived similarity. Dots (left) represent data points of individual participants, violins (right) distribution
within conditions}
  \label{fig:similarity}
\end{figure}

Hypothesis 1 examined the effect of the degree of avatar-person similarity manipulation (condition) on perceived similarity. The ANOVA revealed a significant main effect for condition, \textit{F}(1.87,59.72) = 10.13, \textit{p} \textless .001, $\eta^2$ = .11. 
Subsequent comparisons did not show a significant effect between the 0\%- (\textit{M} = 3.34, \textit{SD} = 1.26) and the 50\%-condition (\textit{M} = 3.18, \textit{SD} = 1.21), \textit{t}(32) = 1.36, \textit{p} = .183, as well as the CG, \textit{t}(32) = 3.46, \textit{p} = .008 and between the 100\%-condition and the CG, \textit{t}(32) = 2.14, \textit{p} = .08.
All other comparisons showed a significant difference, showing a steady decline in similarity with increasing similarity manipulation (0\% = highest - CG = lowest). This included significant effects between the 0\%- and the 100\%-condition (\textit{M} = 2.75, \textit{SD} = 0.85), \textit{t}(32) = 3.17, \textit{p} \textless .013, as well as the CG (\textit{M} = 2.44, \textit{SD} = 0.73), \textit{t}(32) = 3.83, \textit{p} = .003. Further, between the 50\%- and 100\%-condition, \textit{t}(32) = 2.54, \textit{p} = .049.
Thus, H1 was partially confirmed: Manipulation of avatar-person similarity influences the perceived similarity, except for the first two degrees of manipulation (see Fig. \ref{fig:similarity}).

\subsubsection{Explicit Identification}
\begin{figure}[h]
\centering
  \includegraphics[width=0.5\textwidth]{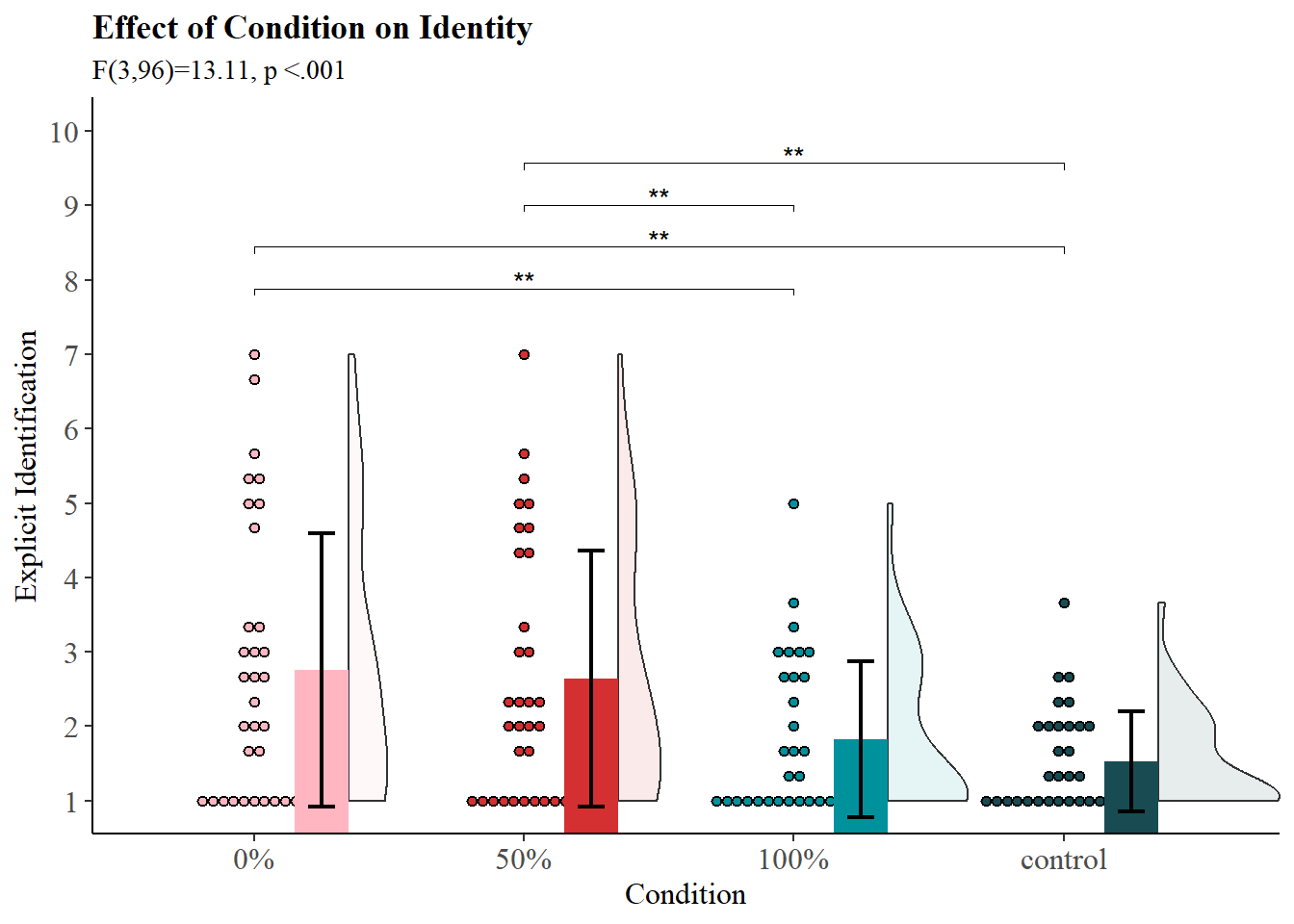}
  \caption{Effects of similarity manipulation on explicit identification. Dots (left) represent data points of individual participants, violins (right) distribution within conditions}
  \label{fig:Identification}
\end{figure}

Hypothesis 2 examined the effect of the degree of avatar-person similarity manipulation (condition) on explicit identification. The ANOVA revealed a significant main effect for condition, \textit{F}(2.03,65.01) = 13.11, \textit{p} \textless .001, $\eta^2$ = .13  
Subsequent comparisons did not show a significant effect between the 0\%- (\textit{M} = 2.76, \textit{SD} = 1.84) and the 50\%-condititon (\textit{M} = 2.64, \textit{SD} = 1.72), \textit{t}(32) = 0.64, \textit{p} = .527, or between the 100\%- (\textit{M} = 1.82, \textit{SD} = 1.05) and the CG (\textit{M} = 1.53, \textit{SD} = 0.67), \textit{t}(32) = 1.68, \textit{p} = .206.
All other comparisons showed a significant difference, showing a decline in explicit identification with increasing similarity manipulation. This included significant effects between the 0\%- and the 100\%, \textit{t}(32) = 3.71, \textit{p} = .002, as well as the CG, \textit{t}(32) = 4.05, \textit{p} = .002. Further, between the 50\%- and 100\%, \textit{t}(32) = 4.03, \textit{p} = .002, as well as the CG, \textit{t}(32) = 4.16, \textit{p} = .001.
Thus, H2 was also partially confirmed: Manipulation of avatar-person similarity influences also explicit identification, except for the first and last two degrees of manipulation (see Fig. \ref{fig:Identification}).

\subsubsection{Implicit Identification (Affinity)}
Hypothesis 3 examined the effect of the degree of avatar-person similarity manipulation (condition) on affinity. The ANOVA revealed a significant main effect for condition, \textit{F}(2.2,70.44) = 7.66, \textit{p} \textless .001 , $\eta^2$ = .05.
Subsequent comparisons again did neither find a significant effect between the 0\%- (\textit{M} = 3.22, \textit{SD} = 1.77) and the 50\%-group (\textit{M} = 3.08, \textit{SD} = 1.72), \textit{t}(32) = 0.95, \textit{p} = .694, nor between the 100\%- (\textit{M} = 2.52, \textit{SD} = 1.51) and the CG (\textit{M} = 2.37, \textit{SD} = 1.3), \textit{t}(32) = 0.74, \textit{p} = .694.
However, all of the other group comparisons showed a significant difference, showing a decline in explicit identification with increasing similarity manipulation. 

\begin{figure}[h]
\centering
  \includegraphics[width=0.5\textwidth]{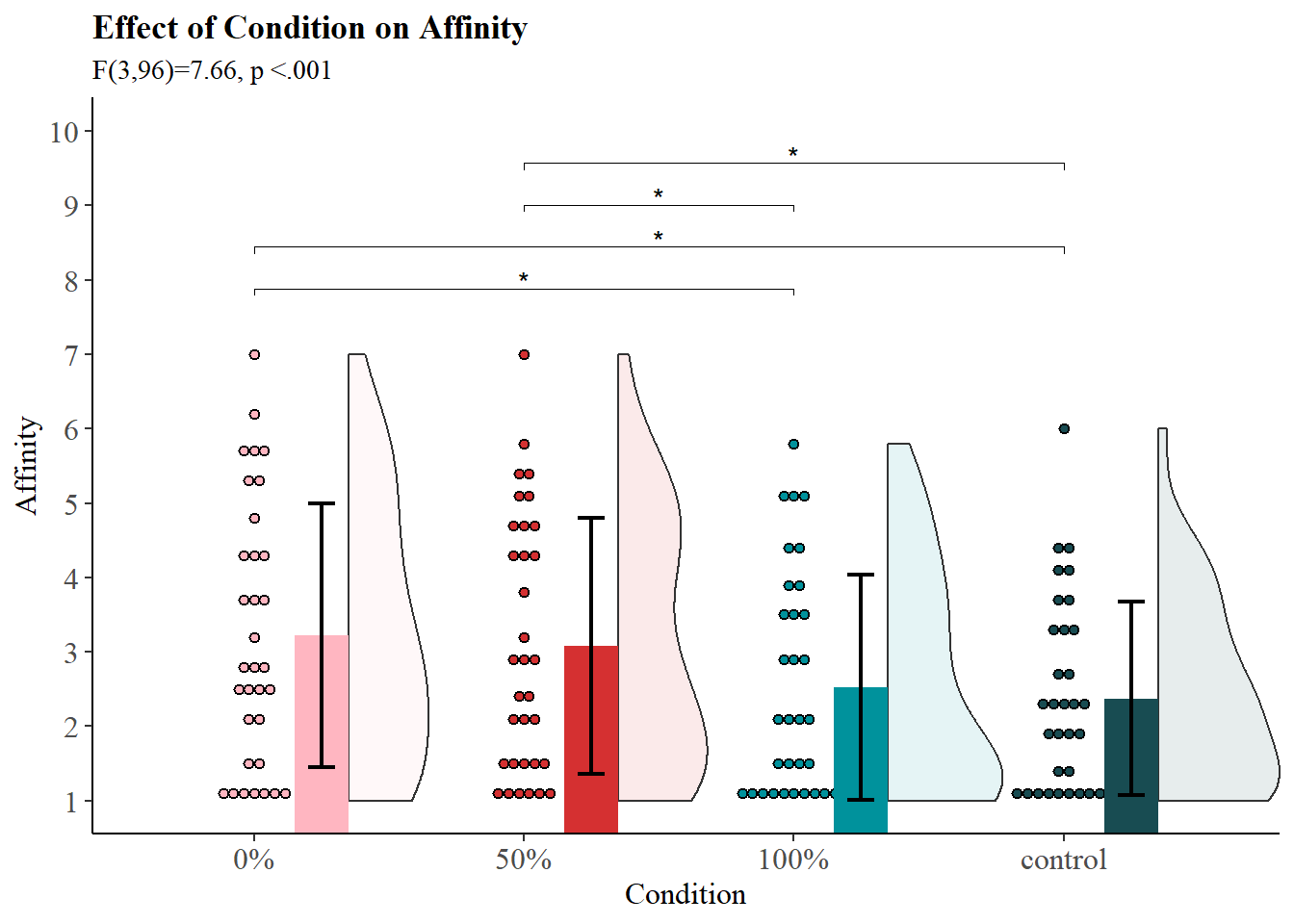}
  \caption{Effects of similarity manipulation on affinity. Dots (left) represent data points of individual participants, violins (right) distribution
within conditions}
  \label{fig:Affinity}
\end{figure}

This included significant effects between the 0\%- and the 100\%, \textit{t}(32) = 3.30, \textit{p} = .014, as well as the CG, \textit{t}(32) = 3.32, \textit{p} = .014. Further, between the 50\%- and 100\%, \textit{t}(32) = 3.17, \textit{p} = .014, as well as the CG, \textit{t}(32) = 2.77, \textit{p} = .028.
Thus, H2 was again partially confirmed: Manipulation of avatar-person similarity influences also explicit identification, except for the first and last two degrees of manipulation (see Fig. \ref{fig:Affinity}).

\subsubsection{Exploratory in-depth analysis}
In order to gather learnings for follow-up studies and future research, we performed an exploratory in-depth analysis. As a first step, correlations between the dependent variables and the demographic variables were examined. Results can be found in Table \ref{correlations}. In an additional analysis of covariances, we examined whether the variables realism, gender and previous experience with an avatar played a role. The analysis indicated an influence of the covariates in a type II ANCOVA, that did not show anymore when performing the type III variant, but no significant effect was found overall. The exact parameters, R scripts and results of this analysis can be found in the additional online material on OSF.

\begin{table}[htbp]
\centering
\caption{Means, standard deviations, and correlations with confidence intervals}
\label{correlations}
\begin{tabular}{llcccccccc}
\toprule
&Variable & M & SD & 1 & 2 & 3 & 4 & 5 & 6\\
\midrule
1 & Similarity & 2.93 & 1.08\\
\\
2 & IE & 2.80 & 1.61 & .82$^{**}$\\
& & & & [.76, .87] \\
3 & EE & 2.18 & 1.48 & .84$^{**}$ & .85$^{**}$\\
& & & & [.79, .89] & [.79, .89] \\
4 & Realism & 4.44 & 1.68 & .45$^{**}$ & .51$^{**}$ & .44$^{**}$\\
        & & & & [.31, .58] & [.38, .63] & [.29, .57] \\
5 & Attachment & 2.85 & 0.40 & .08 & .24$^{**}$ & .14 & .26$^{**}$ \\
        & & & & [-.09, .25] & [.08, .40] & [-.03, .30] & [.09, .41] \\
6 & AfT & 5.91 & 1.36 & -.22$^{*}$  & -.29$^{**}$ & -.22$^{*}$ & -.04 & .09\\
& & & & [-.37, -.05] & [-.44, -.12] & [-.38, -.05] & [-.21, .13] &[-.09, .25] \\
7 & Experience & 1.76 & 0.43 & .12 & .10 & -.01 & .19$^{*}$ & .22$^{*}$ & -.04 \\
& & & & [-.05, .29] & [-.08, .26] & [-.18, .17] & [.02, .35] & [.05, .38] & [-.21, .13] \\
\bottomrule
\end{tabular}
\medskip
\begin{flushleft}
\footnotesize{Note. M and SD are used to represent mean and standard deviation, respectively. EE = Explicit Identification. IE = Implicit identification/Affinity. Attachment = Attachment style. AfT = Affinity for Technology. Experience = Previous experience with avatars. Values in square brackets indicate the 95\% confidence interval. The confidence interval is a plausible range of population correlations that could have caused the sample correlation \cite{cumming2014new}. * indicates $p < .05$. ** indicates $p < .01$.}
\end{flushleft}
\end{table}

\section{Discussion}
This user study investigated the effect of manipulating specific facial features of virtual avatars by degrees of visual similarity to the participants on perceived similarity of the avatar, as well as two antecedents of self-awareness: explicit and implicit identification.
The process of developing the methodology for avatar similarity variation involved the investigation of tools and solutions, the testing of their functionalities, and research on the theoretical background of aspects in which the methodology could be applied (i.e. visual similarity, affinity, identification, etc).
A similarity scale was defined, formalizing three degrees of similarity variation (high similarity - 0\%, medium similarity-50\%, low similarity-100\%). 
A user study was set up and performed to validate the application of the proposed methodology and test the effects of systematic similarity variation on the antecedents of self-awareness. 
Spefically, we hypothesized that a higher degree of visual similarity would lead to higher perceived similarity as well as explicit and implicit identification/affinity. The results of the study generally confirm these hypotheses.
		
\subsection{User Study: Hypothesis Discussion}
Manipulation check:
All degrees except the 100\%-manipulation were perceived as realistic. This lets us assume that all effects, especially the ones between the 0\%-, 50\%- and CG, can be attributed to the manipulation rather than an unnatural distortion of the participants' faces.

H1:
Manipulating visual avatar-person similarity using the developed methodology indeed showed an effect on the similarity perceived by the user in the expected direction (confirming H1). This also serves as an additional manipulation check for the following analyses. Single comparisons revealed, that there was a steady decline in perceived similarity between degrees of manipulation, indicating that the approach of adjusting specific facial features to manipulate the degree of similarity is suitable. 
This result provides an argument against generic or averaged avatars.
The difference between 0\%- and 50\%-manipulation, although objectively different, was not perceived as dissimilar. This effect could be explained if we assume that the manipulation was still perceived too similar in the two conditions and gives future researchers and practitioners a bit of an error margin to work with, in terms of manipulating visual similarity.

H2:
Avatar-person similarity also showed an effect on the explicit identification by the user in the expected direction (confirming H2). Thus, the participants tended to identify more with an avatar, the more it was objectively similar to them. While not particularly surprising, this result has important implications, given that identification is one of the key antecedents in eliciting self-awareness.
Examining the distributions of answers between the four conditions closely, however, in the 0\% and 50\%-condition, a wide range of answers was observable, compared to the other two conditions, where answers largely concentrated on the lower end of the scale.
Single comparisons revealed that there was a steady decline in explicit identification with decreasing similarity. 
Taken together, these detailed analyses deliver a strong argument against generic, non-personalized avatars whenever explicit identification with the avatar and self-awareness elicitation are goals in the study.
Here too, the difference between 0\%- and 50\%-manipulation did not have an effect on explicit identification by the user, and neither did the difference between 100\%-manipulation and the CG. This effect could be explained assuming, that for a difference in identification, a small change in the avatar seems not enough, while in the 100\%-condition, the participants apparently did not recognize themselves anymore (as further indicated by the significant effect between 50\%- and 100\%-, but not 100\%-manipulation and CG). An interaction with the missing realism could be claimed here. However, the missing effect in the CG, which was perceived as realistic, speaks against that.  
We can conclude that, in terms of explicit identification, the right amount of similarity manipulation is important.

H3:
Lastly, manipulating visual avatar-person similarity indeed showed an effect on implicit affective identification (affinity) in the expected direction. Apart from merely identifying with the avatar, participants also rather created an emotional bond - an affinity - towards avatars that are more similar to them. 
Single comparisons revealed the same pattern of effects between the conditions as in explicit identification, again indicating, that, whenever affinity or self-awareness are to be elicited, using a generic avatar could be counterproductive. 
Similarly, the difference between 0\%- and 50\%-manipulation did not produce an effect on affinity, neither did the difference between 100\%-condition and the CG. 
Again, a small change in avatar-person similarity did not seem to affect the affection towards the avatar, while in the 100\%-condition, the participants apparently did not feel an emotional affinity anymore (as indicated by the sig effect between 50 and 100 but not 100 and CG). 

Exploratory in-depth analysis:
The dependent variables showed significant correlations between each other as well as to possible demographic covariates such as realism, attachment style, affinity for technology and previous experience with avatars. The former result was expected given the conceptual closeness of the given constructs. Therefore, our results do show that the concepts are distinct but not disjoint. In order to differentiate the relevant subtle differences between them, a fine-grained differentiation as presented in this study is necessary. 
Naturally, in a highly-individualized setting like personal avatars, individual differences and attitudes of participants may influence results. While we could find indications that the examined covariates do play a role (comparison of type II vs. type III ANCOVA, see results in OSF), no significant effects could be reported. This might be due to the fact that the manipulation and sample size of this particular study were not explicitly set-up for this kind of analysis. 
However, reporting these indications is still relevant, for them to be taken into account for future studies and implementations to avoid "one-size-fits-all" applications.
		
\subsection{General Discussion}
This research presents a novel approach using avatars in virtual environments as a tool to systematically manipulate and test the effects of unavoidably induced self-awareness.
One of the most challenging aspects of this research relies on developing and implementing an approach to creating humanoid avatars that accurately represent participants (RQ 1). 
The detailed reproducible step-by-step description of the methodology presented in this article was aimed at introducing a generalizable open-access approach of creating visually similar avatars. It allows by design that avatars present dynamic features (being able to show facial expressions through animations) and that such avatars could be easily manipulated to create variations of them (degrees of similarity between the avatar and the user).
One of the major advantages of the approach chosen in this work, is that it was all based on openly accessible tools, so the methodology could be easily adapted for future research by other research teams.
The user study suggests that the proposed methodology including the definition of the manipulation of relevant facial features can be applied to vary visual avatar-person similarity through facial feature manipulation (RQ 2). 
The presented results are in line with past research indicating that avatar-person identification corresponds with perceived visual similarity. 
Further, as this research aims to examine the effect of visual similarity on the elicitation of self-awareness and its antecedents on a process level for its use in interactive virtual trainings, we aimed for a more fine-grained examination of visual similarity manipulation beyond a simple distinction between personalized vs. non-personalized. 
Compared to previous studies, this article examined in closer detail how avatar-person similarity is processed by users and the effects it produces. 
Introducing this custom methodology revealed that perceived similarity as well as explicit and implicit affective identification with virtual avatars are both dependent on the degree of visual avatar-person similarity and highlighted, that generic avatars are insufficient for manipulations that depend on them, like self-awareness. 
As visual avatar-person similarity is correlated with explicit identification \cite{vasalou2007constructing} and affinity (i.e., liking persons perceived as similar) which both were found to predict self-awareness \cite{hooi2012being, hooi2014avatar}, next steps in future research might involve the identification and testing of other aspects of similarity, such as behavioural contingency, to increase support self-awareness.
Given that self-awareness and consequently, self-compassion are key ingredients in reflection and emotion regulation \cite{gyurak2011explicit}, the effects of interactive virtual trainings such as \cite{bhuvaneshwara2023mithos}, where teachers interact with interactive virtual agents (students), could be amplified significantly by personalizing teacher avatars to increase their self-awareness.
These findings also have important implications for the design of virtual environments in the area of interactive virtual training environments in general, such as online educational games, simulations and virtual worlds, where the use of avatars may be beneficial for training situations. 
This is especially important for scenarios where constructs related to perceptions of the self, like self-awareness, self-compassion (and its subcomponents like mindfulness), but also presence and immersion are aimed at \cite{chandrasiri2020virtual}.
Similarly, in light of our in-depth analysis, the application of this method to virtual training environments should take into account that in personalized settings like these, individual preconditions of the participants should be considered. For example, given the correlation between attachment style and the dependent variables in this study, a personalized style of animations as proposed in \cite{reinwarth2023look} might be an initial step in this direction.

On the practical side, these research findings are significant in light of the fact that many applications from computer games to educational tools allow players to create and customize avatars. A special advantage of the used methodology for the scenario of the research is that it allows exporting the 3D avatar to animation tools, which can be used to personalize facial animations including eye tracking.  Most importantly, this research paves the way to automatic personalized behavior feedback \cite{bhuvaneshwara2023mithos, alves2023visual}.
Further, the methodology was designed in a way that can generalize to improvements in technology for avatar creation.
With technology evermore advancing, virtual representation is already starting to become a key ingredient in virtual reality such as e.g. the Metaverse \cite{lawrence2022state, kim2023avatar} or mixed-reality applications e.g. with the Apple Vision Pro platform \cite{zhang2023apple}. The present study shows, how much of a difference for self-representation and immersion personalized avatars could have. Future work is needed to examine this further.
		
\subsection{Limitations and Future Work}
The technical methodology for creating avatar variations was applied to a study that used animated facial images (GIF) of avatars and it is possible that the results may differ from studies that require the use of dynamic avatars containing the whole body. Equivalently, further research with dynamic avatars is needed to validate the findings and explore the full potential of avatars as a means of increasing self-awareness. Controlling for movement similarity must be considered in that case. 
Regarding the avatar generation, the sample size of the study was relatively small, and further research is needed to replicate the results of the methodology. For both of these limitations, a larger follow-up study is already running at this moment.
Another limitation is in regard to the limitations of the solutions that were used to create avatars and their variations. Some of the visual characteristics (e.g. hair, beard, eyebrows format) have a limited amount of pre-defined choices that might not match some participants' features. 
Further, it might be the case that when the low similarity condition was presented first in the study survey, it set the scene for the perception of all avatars, and caused the participant to perceive them all as very different from themselves. However, since the order of presentation of the conditions was randomized, a biasing of the results through this special case is rather unlikely.
More data needs to be collected to validate that limitation. 
Future research applying this methodology for manipulating perceived similarity, explicit identification or affinity should take into account that this manipulation has to be made large enough in order to actually make a difference.
Finally, future research should consider applied research scenarios to validate the methodology for other effects measurements when considering avatar-person similarity. As such, a series of follow-up studies is currently in the phase of data collection, applying personalized avatars created through the presented methodology in an interactive mixed-reality classroom training environment \cite{bhuvaneshwara2023mithos}. This will show whether the effects found in this study will replicate in an applied, dynamic and interactive setting.

\section{Conclusion}

This article presents a theory-based methodology to manipulate the facial features of virtual avatars by degrees of visual similarity with the aim of affecting perceived similarity to the user (RQ 1). We evaluated this methodology and showed the effect of manipulating the degree of visual avatar-person similarity on the antecedents for increasing self-awareness with avatars, namely, perceived similarity, explicit identification and implicit identification/affinity (RQ 2). 
 
In this work, we aimed for a  fine-grained examination of visual similarity manipulation beyond a simple distinction between personalized vs. non-personalized avatars. Our results underline that in scenarios in which constructs related to perceptions of the self, like self-awareness, self-compassion (and mindfulness), but also presence and immersion are aimed at, generic avatars might not be enough providing a foundation for future work in this area.
Therefore, it is crucial for future HCI research to consider more nuanced approaches to visual similarity manipulation in these scenarios to achieve better outcomes. This work contributes to the research on increasing awareness and paves the way for the automatic generation of similar avatars. It is important to note that although the results of the study showed that the highest, albeit not perfect, avatar visual similarity is conducive to antecedents of awareness, it is still possible that this effect depends on individual differences, e. g.  trait self-consciousness \cite{chehayeb2021individual}. Automatically manipulating and adapting the degree of avatar visual similarity, can be used to support self-awareness in virtual environments, while taking into account individual differences. Additionally,  this research enables to variably manipulate the behaviour of the personalised avatar for self-awareness in immersive virtual environments, for instance to train for further behavioural change.





\section*{Acknowledgements}

{\label{687807}}

MITHOS: This work was funded by the German Federal Ministry of Education and Research: 16SV8687.
We thank Joanna Mauer, Tobias Lang, Anna Lea Reinwarth, Janet Wessler, Tanja Schneeberger and Benedikt Wirth for their help with this study.

\selectlanguage{english}
\FloatBarrier

\bibliographystyle{acm}
\bibliography{sample-base}




\end{document}

\documentclass{article}
\usepackage{graphicx} 

\title{Similarit_Study_CHB}
\author{bhilpert94 }
\date{December 2023}

\begin{document}

\maketitle

\section{Introduction}

\end{document}